\documentclass[a4paper,conference]{IEEEtran}

\usepackage[utf8]{inputenc} 
\usepackage[english]{babel}

\def\mytitle{Stable smoothed particle magnetohydrodynamics in very steep density gradients}
\def\myauthor{Benjamin T. Lewis \& Matthew R. Bate}

\def\rmn{\mathrm}

\usepackage{color,soul} 
\def\diff{} 
\newcommand{\ignore}[1]{}

\usepackage{cite}      

\usepackage{graphicx}  

\usepackage{subfigure} 

\usepackage{amsmath}   
\usepackage{amssymb} 

\usepackage{aas_macros}
\usepackage{textcomp}
\usepackage{gensymb}
\usepackage{empheq}
\usepackage{csquotes}
\usepackage[hidelinks, pdfauthor={\myauthor}, pdftitle={\mytitle}, pagebackref=false]{hyperref}
\usepackage{cleveref}

  \crefname{figure}{Fig.}{Figs.}
  \Crefname{figure}{Fig.}{Figs.}
  \crefname{section}{$\S$}{$\S\S$}
  \Crefname{section}{$\S$}{$\S\S$}
  \crefname{equation}{Eqn.}{Eqns.}
  \Crefname{equation}{Eqn.}{Eqns.}

\usepackage{fancyheadings}
\begin{document}

\title{\mytitle }

\author{\IEEEauthorblockN{Benjamin T. Lewis, Matthew R. Bate}
\IEEEauthorblockA{School of Physics and Astronomy\\
University of Exeter\\
Exeter, EX4 4QL\\
blewis@astro.ex.ac.uk}
\and
\IEEEauthorblockN{Joseph J. Monaghan, \& Daniel J. Price}
\IEEEauthorblockA{Monash Centre for Astrophysics\\
Monash University\\
Clayton, Vic 3800\\
}
}


\maketitle

\begin{abstract}
The equations of smoothed particle magnetohydrodynamics (SPMHD), even with the various corrections to instabilities so far proposed, have been observed to be unstable when a very steep density gradient is necessarily combined with a variable smoothing length formalism. Here we consider in more detail the modifications made to the SPMHD equations in LBP2015 that resolve this instability by replacing the smoothing length in the induction and anisotropic force equations with an average \diff{smoothing length} term. We then explore the choice of average used and compare the effects on a test `cylinder-in-a-box' problem and the collapse of a magnetised molecular cloud core. We find that, aside from some benign numerical effects at low resolutions for the quadratic mean, the formalism is robust as to the choice of average but that in complicated models it is essential to apply the average to both equations; in particular, all four averages considered exhibit similar conservation properties. This improved formalism allows for arbitrarily small sink particles and field geometries to be explored, vastly expanding the range of astronomical problems that can be modeled \diff{using} SPMHD.
\end{abstract}

\section{Introduction}
From some of the earliest work on SPH by Gingold and Monaghan \cite{1977MNRAS.181..375G} how to include magnetohydrodynamics in the equations of SPH has been an area of active research. Whilst major progress has been made since then, e.g. the Lagrangian formalism derived by Price and Monaghan \cite{2004MNRAS.348..123P}, the correction to the tensile instability by Borve et al. \cite{2001ApJ...561...82B} (on which there is more discussion below), and the divergence cleaning devised by \cite{2012JCoPh.231.7214T}, SPMHD is known to be unstable in certain astrophysical simulations. A stable formalism of SPMHD is essential to fully study many astrophysical processes, for example the generation of protostellar outflows and collimated jets which remove angular momentum from an accreting protostellar core.

Extreme \diff{gradients in density} are not uncommon in astrophysical simulations, and these can result in \diff{correspondingly very small and very large smoothing lengths} being needed \diff{in a physically small region of the simulation}. In some cases, for example in the pseudo-disc surrounding a protostar the smoothing length at the top of one of these gradients can be so short that the gradient is not properly sampled, \textit{i.e.} \ignore{the neighbour set of a particle in densest region can be almost completely dissimilar to one at the end.}\diff{ a situation can arise where the particle spacing, }$\Delta{}\rho \sim{} \frac{\rho}{h}$\diff{.} In ordinary hydrodynamic SPH (and also when radiative transfer schemes are employed) no deleterious effects are observed since all the equations of SPH either employ both the smoothing length of a particle itself and its neighbours or, in the case of the density, can be solved self-consistently. However, in the most common formalism of magnetohydrodynamics in SPH (see \cref{sec:spmhd}) this is no longer true, causing a violent instability. We find that the instability observed is caused by a very large (and clearly unphysical) magnetic field being produced, which then acts to accelerate particles rapidly away from the density gradient. Importantly, this is not a `divergence explosion' caused by a failure to maintain a solenoidal field - observed values of $\frac{h|\nabla^i B^i|}{|B^i|}$ remain $\lesssim 0.1$ until after the explosion happens.

In \diff{Lewis, Bate, \& Price (submitted) (hereafter `LBP2015')} we replaced the individual smoothing length terms, $h_{\{a,b\}}$, in these equations with an average term, $\bar{h}_{ab} = \frac{1}{2}\left( h_a + h_b \right)$ which has the desirable property that at an extreme gradient $\bar{h}_{ab}$ tends towards $\frac{1}{2}$ of the larger value, ensuring the full gradient is sampled \diff{and that the equations are evaluated over an identical neighbour set}. This modified approach is covered in more detail in \cref{sec:avh}. This modified scheme allows for arbitrarily small sink particles to be employed, and consequently a much larger range of physics to be sampled than hitherto possible. We reject simply imposing a minimum smoothing length for two reasons, firstly that it is essentially impossible to determine \textit{a priori} a correct value to use; and secondly that in practice the minima needed are so large that particle pairing caused by stretching the smoothing kernel becomes an serious issue. In principle, this could be mitigated by the use of a Wendland kernel \cite{2012MNRAS.425.1068D}, but is still wasteful of resolution in regions where the equations are already stable. The use of an average is more nuanced since when $h_a \sim h_b$, $\bar{h}_{ab} \approx h_{\{a,b\}}$ and therefore resolution is not wasted. Additionally, the average can be applied to only the equations which are unstable, preserving full resolution elsewhere.

However, the arithmetic mean is not the only plausible choice of average - \diff{for example the other two Pythagorean means both tend to zero if either of} $h_{\{a,b\}}$ \diff{is zero whilst the arithmetic mean will approach one-half of the non-zero quantity}. In \cref{sec:averages} we consider the potential advantages of other choices of average. In practice, as seen when applied to our test `cylinder-in-a-box' model in \cref{sec:tests}, no substantial difference is observed. Finally, in \cref{sec:protostar} we apply these averages to a sample of models similar to those in LBP2015. 

\section{Smoothed Particle Magnetohydrodynamics}
\label{sec:spmhd}
We are solving the equations of ideal magnetohydrodynamics with a gravitational term using the SPMHD method initially presented in \cite{2004MNRAS.348..123P}. We define the MHD stress tensor as 
\begin{equation}
S^{ij} = -P \delta^{ij} + \frac{1}{\mu_0} \left( B^{i}B^{j} - \frac{1}{2}\delta^{ij}B^2 \right) \text{~,}
\end{equation} 
which can be usefully separated into an isotropic `pressure' component,
\begin{equation}
S^{ij}|_{\rmn{iso}} = - \left( P + \frac{1}{2}\frac{1}{\mu_0}B^2 \right) \delta^{ij} \text{~,}
\end{equation}
and an anisotropic `tension' component,
\begin{equation}
S^{ij}|_{\rmn{anis}} = \frac{1}{\mu_0} B^{i}B^{j} \text{~.}
\end{equation}
We write the total derivative as
\begin{equation}
\frac{\rmn{d}~}{\rmn{d}t} = \frac{\partial{}~}{\partial{}t} + v^{i}\nabla^{i} \text{~,}
\end{equation}
and adopt Einstein's convention so that repeated indicies imply summation. Therefore we can write the equations of ideal MHD as
\begin{equation}
\frac{\rmn{d}~}{\rmn{d}t}\rho{} = -\rho{}\nabla{}^{i}v^{i} ~\text{,}
\end{equation}
\begin{equation}
\frac{\rmn{d}~}{\rmn{d}t} v^{i} = \frac{1}{\rho} S^{ij} - \nabla^{i}\phi{}
\label{eqn:mom_analyitc}
\end{equation}
\begin{equation}
\frac{\rmn{d}~}{\rmn{d}t} B^{i} = \left(B^{j}\nabla^{j} \right)v^{i} - B^{i} \left(\nabla^{j} v^{j} \right) ~\text{,}
\end{equation}
\begin{equation}
\nabla^{2}\phi{} = 4\pi{}G\rho{} ~\text{,}
\end{equation}
where the other symbols have their usual meanings.

In addition to the discretization in \cite{2004MNRAS.348..123P,2005MNRAS.364..384P}, we add artificial viscosity and resistivity terms, as a result the final equations are not exactly `ideal' MHD. We use the Riemann solver based \diff{artificial} dissipation terms of \cite{1997JCoPh.136..298M} with spatial and temporally varying switches to reduce the dissipation to the minimum necessary to maintain numerical stability. The older \cite{1997JCoPh.136...41M} switch is used for the \diff{artificial} viscosity term, with $\alpha_{\rmn{AV}} \in [0.1, 1.0]$ but the newer \cite{2013MNRAS.436.2810T} switch for \diff{artificial} resistivity (which is observed to exhibit greater stability in protostellar collapse simulations) with $\alpha_{\rmn{B}} \in [0.0, 1.0]$. The \diff{gravitational} forces are solved using a binary tree and softened using the SPH smoothing kernel \cite{2007MNRAS.374.1347P}.

All magnetic fields found in nature are solenoidal, and therefore 
\begin{equation}
\nabla{}\cdot{}\mathbf{B} = 0 \text{~.}
\end{equation}
This constraint is only present in the equations of SPMHD as an initial condition; due to round-off error  
\begin{equation}
\frac{\rmn{d}~}{\rmn{d}t} \nabla{}\cdot{}\mathbf{B} \neq 0
\end{equation}
in general and as a result, the magnetic field will rapidly become non-solenoidal and no longer correct. This will \textit{inter alia} produce an unphysical force along the magnetic field lines \cite{1995JCoPh.116..123S} since the $B^{i}\left(\nabla^{j}B^{j}\right)$ term in the anisotropic part of the SPMHD momentum equation
\begin{align}
\begin{split}
\frac{\rmn{d}~}{\rmn{d}t} v^{i}|_{\rm{anis}} & = - \frac{1}{\rho{}}\nabla^{j}S^{ij}|_{\rmn{anis}} = - \frac{1}{\rho{}}\frac{1}{\mu_{0}}\nabla^{j} B^{i}B^{j} \\
& = -\frac{1}{\rho{}}\frac{1}{\mu_{0}} \left[ \left(B^{j}\nabla^{j}\right)B^{i} + B^{i}\left(\nabla^{j}B^{j}\right)\right] \text{~,}
\end{split}
\end{align}
will no longer be zero.
An effective correction to this is to subtract a source term exactly equal to this divergence \cite{2001ApJ...561...82B}, which produces an SPMHD momentum equation (neglecting the isotropic pressure terms which are unchanged) that depends only on $h_b$, \textit{i.e.} the smoothing length of each of the particles neighbours, \textit{viz.}
\begin{equation}
\frac{\rmn{d}~}{\rmn{d}t} v^{i}_a|_{\rmn{anis}} = \frac{1}{\mu{}_{0}} \sum^{N}_{b} \frac{m_{b}}{\Omega_{b}\rho^{2}_{b}}\left(B^{i}_{b} - B^{i}_{a} \right)B^{j}_{b}\nabla{}^{j}_{a}W_{ab}\left(h_{b}\right) \text{~.}
\label{eqn:spmhdaniso}
\end{equation}
where $W_{ab}$ is the SPH smoothing kernel and 
\begin{align}
\begin{split}
\Omega_{a} &= 1 - \frac{\partial{}h_{a}}{\partial{}\rho_{a}} \sum^{N}_{b} \frac{\partial{}W_{ab}\left( h_a\right) }{\partial{}h_{a}}\\
&= 1 + \frac{h_{a}}{\nu{}\rho_{a}}\sum^{N}_{b}\frac{\partial{}W_{ab} \left( h_{a} \right) }{\partial{}h_{a}}\text{~.}
\label{eqn:omega}
\end{split}
\end{align}

In comparison, the SPMHD induction equation depends only upon $h_a$, \textit{i.e.} the particle's own smoothing length, 
\begin{equation}
\frac{\rmn{d}~}{\rmn{d}t} \left( \frac{B^{i}_{a}}{\rho_{a}} \right) = -\frac{1}{\Omega_{a}\rho^{2}_{a}} \sum^{N}_{b} m_{b} \left( v^{i}_{a} - v^{i}_{b} \right) B^{j}_{a} \nabla^{j}_{a}W_{ab}\left(h_{a}
\right) \text{.}
\label{eqn:spmhdind}
\end{equation}
When coupled with a form of magnetic divergence cleaning this formalism is remarkably robust. For this work, we use the constrained hyperbolic divergence cleaning derived in \cite{2012JCoPh.231.7214T}, where a new scalar field, $\psi$, is coupled to the magnetic field such that
\begin{equation}
\frac{\rmn{d}~}{\rmn{d}t} B^{i}|_{\rmn{clean}} = - \nabla^{i}\psi
\end{equation}
and where $\psi$ is evolved by
\begin{equation}
\frac{\rmn{d}~}{\rmn{d}t} \psi = - c^2_c\nabla^{i}B^{i} - \frac{\psi}{\tau} - \frac{1}{2}\psi\left( \nabla^{i}v^{i} \right)
\end{equation}
where the timescale for damping,
\begin{equation}
\tau = \frac{h}{\sigma{}c_c} \text{~.}
\end{equation}
A value of $\sigma = 0.8$ is used as recommended by \cite{2012JCoPh.231.7214T} to critically damp the cleaning wave over a small number of smoothing lengths.

\section{The `Average h' Method}
\label{sec:avh}
However, when a very large density gradient is present -- \textit{e.g.} in the collapse of a molecular cloud core -- this \diff{method} rapidly becomes unstable. If a variable smoothing length regime is employed (which is essential in any calculation of this nature) where $h$ is a function of $\rho$ for example, that given in \cite{2004MNRAS.348..139P}) where
\begin{equation}
h = \eta\left(\frac{m}{\rho}\right)^{\frac{1}{\nu}}
\end{equation}
(where $\nu = 3$ is the number of spatial dimensions and $\eta$ is a parameter controlling the \diff{typical} number of particles in the smoothing sphere) then when $\rho_a \gg \rho_b \rightarrow h_b \gg h_a$ and \textit{vice versa} it is possible for a particle to have the anisotropic component of its magnetic force evaluated over a very large set of neighbours and the induction equation evaluated over very few. \diff{Consequently, particles will interact with each other for one equation, but not the other, and an inconsistent estimate of both the force and magnetic induction will be calculated.} Analysis of previous protostellar collapse simulations indicate that, for a cubic B-spline kernel with $\eta = 1.2$ and therefore $\approx 53$ neighbours on average, ratios of  greater than 100:4 neighbours are possible. It is worth noting here that whilst $\eta$ controls the neighbour count, it does not guarantee that any, or every, individual particle will have exactly $N_{\rmn{ngh}}$ neighbours unless the average density profile is flat.

The result is a violent instability that disrupts the simulation. In LBP2015 we replaced the $h_{\left\{a,b\right\}}$ terms in \cref{eqn:spmhdaniso,eqn:spmhdind} with an average term,
\begin{equation}
 \bar{h}_{ab} = \frac{1}{2}\left( h_{a} + h_{b} \right) \text{~,}
 \label{eqn:arithmean}
\end{equation}
to prevent this instability and were consequently able to follow the collapse of a protostar much further than previously possible. Whilst this does require the removal of the $\Omega_{\{a,b\}}$ \cref{eqn:omega} terms, in practice the additional error produced is negligible. We therefore obtain in place of \cref{eqn:spmhdaniso} and \cref{eqn:spmhdind},
\begin{equation}
\frac{\rmn{d}~}{\rmn{d}t} v^{i}_a|_{\rmn{anis}} = \frac{1}{\mu{}_{0}} \sum^{N}_{b} \frac{m_{b}}{\rho^{2}_{b}}\left(B^{i}_{b} - B^{i}_{a} \right)B^{j}_{b}\nabla{}^{j}_{a}W_{ab}\left(\bar{h}_{ab}\right) \text{~,}
\label{eqn:spmhdanisohav}
\end{equation}
\begin{equation}
\frac{\rmn{d}~}{\rmn{d}t} \left( \frac{B^{i}_{a}}{\rho_{a}} \right) = -\frac{1}{\rho^{2}_{a}} \sum^{N}_{b} m_{b} \left( v^{i}_{a} - v^{i}_{b} \right) B^{j}_{a} \nabla^{j}_{a}W_{ab}\left(\bar{h}_{ab}
\right) \text{.}
\label{eqn:spmhdindhav}
\end{equation}

\section{Comparison of Averages}
\label{sec:averages}
The arithmetic mean used in LBP2015 is not the only plausible average. Consequently, we considered the effect of choosing a different average. Whilst there are limitless potentialy viable averages, we only consider the three Pythagorean means, \textit{viz.} the arithmetic mean in \cref{eqn:arithmean} (hereafter `A'), the geometric mean (`G') defined as
\begin{equation}
 \bar{h}_{ab} = \sqrt{h_{a}h_{b}} \text{~,}
 \label{eqn:geommean}
\end{equation}
and the harmonic mean (`H') defined as
\begin{equation}
 \bar{h}_{ab} = \frac{2h_{a}h_{b}}{h_{a} + h_{b}} \text{~,}
 \label{eqn:harmmean}
\end{equation}
in addition to the quadratic mean (`Q') defined as 
\begin{equation}
 \bar{h}_{ab} = \sqrt{\frac{1}{2} \left( h_{a}^2 + h_{b}^2 \right) } \text{~.}
 \label{eqn:quadmean}
\end{equation}

All these options take the same value when $h_a = h_b$ but exhibit significant differences, as seen in \cref{tbl:meanlimits}, in the limiting case where $h_a \gg h_b$ or \textit{vice versa}. In particular, whilst A and Q are non-zero in the limit $h_{\{a,b\}} \rightarrow 0$, G and H are not. \diff{It can also be shown that for all values of} $h_a$ \diff{and} $h_b$\diff{,} $A \geq G \geq H$\diff{.} At the other extreme, the differing growth rates of the various means can be seen when $h_a = 10h_b$, where Q is nearly 4 times larger than H. In essence, this affects how much of the gradient each average samples, not only in the limit when the difference between $h_a$ and $h_b$ can be greater a factor of 10, but also when the gradient is shallower. 

\begin{table}
\renewcommand{\arraystretch}{1.3}
\caption{Behaviour of the four means detailed \diff{in }\cref{sec:averages} in limiting cases}
\label{tbl:meanlimits}
\begin{center}
\begin{tabular}{|c|c|c|c|c|}
\hline
\bf{Mean}    & \bf{$h_a \rightarrow 0$} & \bf{$h_a = 2h_b$} & \bf{$h_a = 10h_b$} & \bf{$h_a \rightarrow \infty$} \\
\hline
A            & $\frac{1}{2}h_b$         & $\frac{3}{2}h_b$  & $\frac{11}{2}h_b$ & $\infty$                     \\
\hline
G           & 0                        & $\sqrt{2}h_b$     & $\sqrt{10}h_b$ & $\infty$                      \\
\hline
H           & 0                        & $\frac{4}{3}h_b$ & $\frac{20}{11}h_b$ & $\infty$                       \\
\hline
Q            & $\frac{1}{\sqrt{2}}h_b$ & $\sqrt{5}h_b$ & $\sqrt{\frac{101}{2}}h_b$ & $\infty$                     \\
\hline
\end{tabular}
\end{center}
\end{table}

\section{Numerical Tests}
\label{sec:tests}

\begin{figure*}
\centering{}
\includegraphics{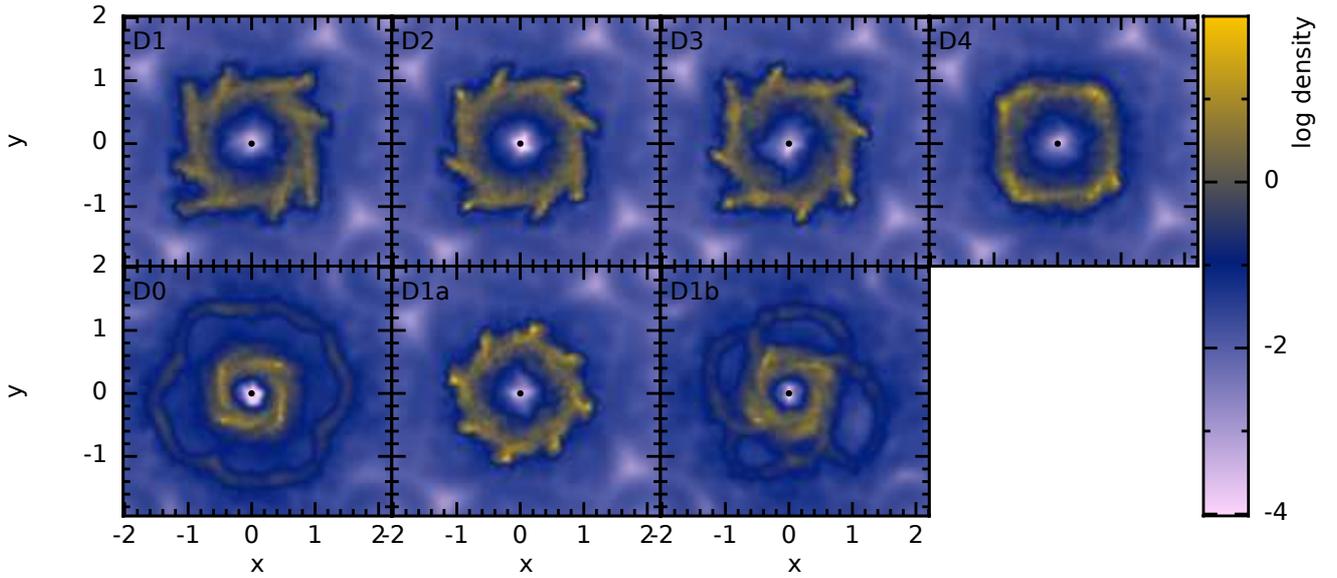}

\caption{Density cuts at $t = 4.5$ of an isothermal test cylinder initially in $r^{-2}$ differential rotation with sink particle providing a central potential equivalent to ten times the mass of the cylinder material. The top row (models D1-4) are the result of using the four averages described in \cref{sec:averages}, all of which remain stable and conservative. The bottom row shows the unmodified equations (D0) and models where only the induction equation \cref{eqn:spmhdind} (D1a) or anisotropic momentum equation \cref{eqn:spmhdaniso} (D1b) are modified. A characteristic unphysical bubble can be seen in both D0 and D1b.
\label{fig:Multi}}
\end{figure*}

We use the simple isothermal cylinder-in-a-box test from LBP2015 to compare each of these averages and an unmodified code. A cylinder of radius $r_{\rmn{cyl}} = 5$ code units \diff{(units defined such that } $G = 1$ \diff{and} $\mu_0 = 1$\diff{)} with a height-to-radius of $\frac{1}{2}$ (2.5 code units thick) and a central hole of radius $\frac{1}{10}r_{\rmn{cyl}} = 0.5$ code units was placed in a periodic box with a central sink particle to provide a potential equivalent to 10 times the mass of the material in the cylinder. \diff{Sink particles (as detailed more comprehensively in }\cite{1995MNRAS.277..362B}\diff{) are particles that exert no force on the system other than gravity and with an `accretion radius', }$r_{\rmn{acc}}$\diff{, whereby any SPH (\textit{i.e.} gas) particle which passes within }$r_{\rmn{acc}}$\diff{ of the sink particle is eliminated from the simulation and its mass and momentum added to the sink particle. We use a sink rather than a simpler potential well since this will eliminate any particles which fall out of the cylinder and into the centre preventing the timestep from becoming needlessly small -- since we are using an isothermal equation of state, the Courant-limited timestep of particles collecting in a central well is very short compared to those in the pressure and magnetically supported cylinder.} The equation of state is given by $P\left(\rho\right) = \frac{2}{3}u\rho$ with $u$ fixed so that the sound speed was 0.1 code units.
An initial magnetic field aligned with the z-axis was applied to give a plasma $\beta$, \textit{i.e.} the ratio of hydrodynamic and magnetic pressure, of $\beta \approx 8.4$. (This is equivalent to that derived by assuming the cylinder is a sphere of material and using the mass-to-flux equations discussed later with $\mu = 5$). 
The cylinder was then given a $r^{-2}$ differential velocity profile with the initial velocity set to obtain a rotation period of $T = 2$ code units at unit radius.

\diff{We would expect the cylinder material to pile up, forming a high density ring with a steep density gradient, so that material within a unit radius moves outwards (since it is moving faster than the Keplerian velocity) and more distant material spiraling inwards. Additionally, some material will fall out of the cylinder and towards the sink particle due to magnetic and viscous braking effect and the cylinder itself will flatten and become more disc like due to rotational and self-gravitational forces.}

Calculations were then performed using the four means presented above (models D1-4) and additionally with the arithmetic mean but applied to either the induction equation only (D1a) or the anisotropic momentum equation only (D1b). We also performed the same simulation with an unmodified code for comparison (D0)\diff{.}

\begin{figure}
\centering{}
\includegraphics{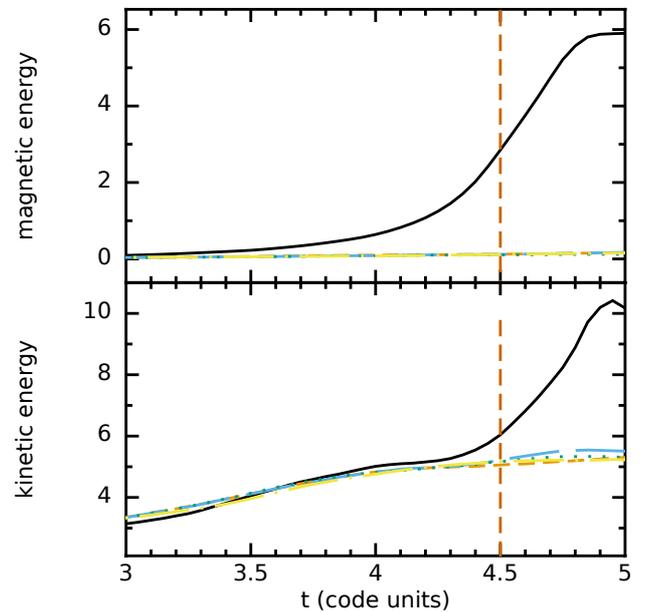}

\caption{The evolution of the total magnetic and kinetic energy for models D0 (solid black line) and D1-4 (dashed orange, long-dashed blue, dotted green, and solid yellow lines respectively). The timestep shown in \cref{fig:Multi} is indicated by a vertical line. All four average $h$ models maintain momentum conservation whilst the unmodified formalism produces a rapidly growing field which ultimately causes an explosion and a consequent increase in kinetic energy \textit{ca.} 1 time unit later.
\label{fig:GrabDisc}}
\end{figure}

In \cref{fig:Multi} we plot the density profile for all seven models just after the explosion has happened in D0. The enhanced stability that using an avergae $h$ formalism provides can be seen in all four models D1-4, however, the D4 model (which is the quadratic mean) has a somewhat dissimilar profile to the three Pythagorean means. The arm-like structures seen in some of the plots are caused by numerical artifacts due to the very low resolution employed. Models D1a and D1b show that, at least in this simple test, it is sufficent to apply the average to the induction equation only; this is expected since the explosion in this instance is clearly driven by an increase in magnetic energy as seen in \cref{fig:GrabDisc}. Even though D4 has evolved somewhat differently, no significant difference in conservation properties is seen between it and the other three average models, as noted in \cref{sec:averages} the quadratic mean will produce larger values for $\bar{h}_{ab}$ in many cases so the differences seen may be due to a somewhat benign loss of effective resolution.

\section{Protostellar Collapse Models}
\label{sec:protostar}

\begin{figure*}
\centering{}
\includegraphics{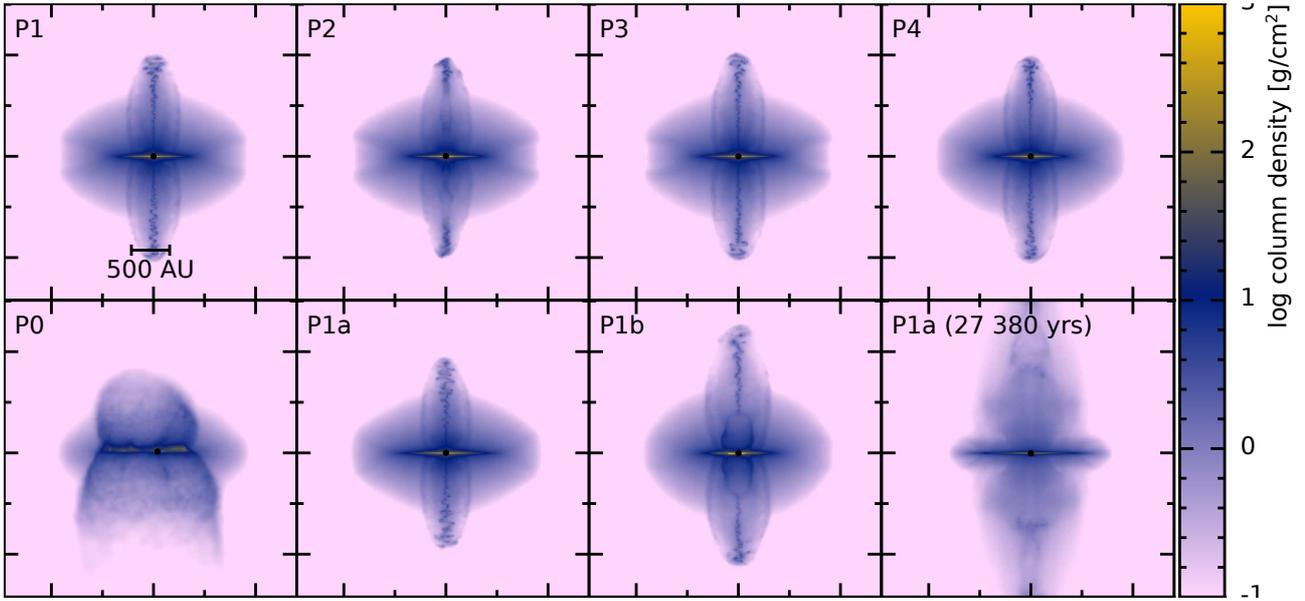}

\caption{Column density plots at $t = 25~670~\rmn{yrs}$ for simulation\diff{s} of \diff{the} collapse of a magnetised molecular cloud core of total mass $1~\rmn{M}_{\odot}$ with a mass-to-flux ratio of $\mu = 5$. The top row (models P1-4) are the four average $h$ models described in \cref{sec:averages}, the stability and essentially identical evolution of which can be clearly seen. Model P0 is an unmodified code, which violently explodes. Model P1a is the same as model P1, \diff{but with the }$h$\diff{-averaging } applied only to the induction equation \cref{eqn:spmhdind}, whilst this is still stable at this timestep, it becomes unstable shortly after (as seen in the far-right plot on the bottom row.) Model P1b has the average applied only to the anisotropic momentum equation \cref{eqn:spmhdaniso} and, whilst significantly improved over P0, still becomes unstable and explodes.
\label{fig:MultiP}}
\end{figure*}

The initial conditions are as detailed in LBP2015. In summary, a $1 \text{M}_{\odot}$ sphere of radius $4\times{}10^{16}~\text{cm}$ surrounded by a warm medium was placed in a periodic box. The sphere was made of \textit{ca.} 1.5 million SPH particles and the warm medium \textit{ca.} 500,000 particles, which is greater than the requirement to resolve the Jeans length in \cite{1997MNRAS.288.1060B}. We set the initial density of the sphere to be $\rho_0 = 7.4\times{}10^{-18} \rmn{g~cm}^{-3}$. The interface between the sphere and medium was in pressure equilibrium but with a 30:1 density contrast, consequently the initial temperature of the sphere is 10 K whilst the surrounding medium is approximately 300 K. A barotropic equation state similar to that in \cite{2008ApJ...676.1088M}, is used, where

\begin{empheq}[left={P = c^{2}_{s} ~ \empheqlbrace}]{equation}
\begin{aligned}
& \rho{}          &&\rho{} \leq \rho_{\rmn{c1}} \\
& \rho_{\rmn{c1}} \left( \frac{\rho{}}{\rho_{\rmn{c1}}} \right)^{\frac{7}{5}} & \hspace{-5pt}\rho_{\rmn{c1}} <~ & \rho{} \leq \rho_{\rmn{c2}} \\
& \rho_{\rmn{c1}} \left( \frac{\rho_{\rmn{c2}}}{\rho_{\rmn{c1}}} \right)^{\frac{7}{5}} \rho_{\rmn{c2}}\left( \frac{\rho{}}{\rho_{\rmn{c2}}} \right)^{\frac{11}{10}} && \rho{} > \rho_{\rmn{c2}}
\end{aligned}
\end{empheq}
to approximate the change in effective $\gamma$ (see \cite{1969MNRAS.145..271L}) as the sphere collapses. The two critical densities are set to $\rho_{\rmn{c1}} = 10^{-14} \rmn{g}~\rmn{cm^{-3}} \approx 10^{3}\rho_0$ and $\rho_{\rmn{c2}} = 10^{-10} \rmn{g}~\rmn{cm^{-3}} \approx 10^{7}\rho_0$.

\begin{figure}
\centering{}
\includegraphics{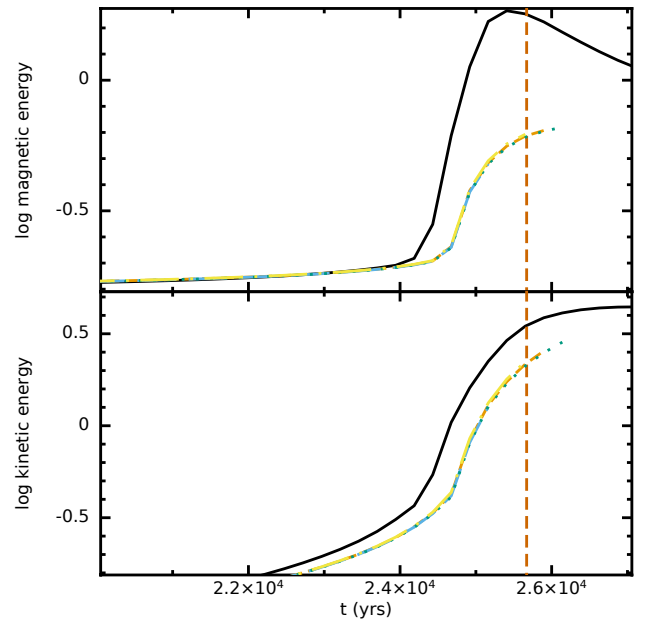}

\caption{The evolution of the total magnetic and kinetic energy for models P0 (solid black line) and P1-4 (dashed orange, long-dashed blue, dotted green, and solid yellow lines respectively). The timestep shown in \cref{fig:MultiP} is indicated by a vertical line. All four average $h$ models approximately maintain energy conservation, however, the P0 model exhibits a sharp increase in both the total magnetic and kinetic energy. Unlike in \cref{fig:GrabDisc}, the increase in magnetic energy does not clearly precede the kinetic energy `knee'. 
\label{fig:GrabCompP}}
\end{figure}

We set the initial magnetic field $B_0$ according to the dimensionless parameter $\mu$, the mass-to-flux ratio of the sphere. This is defined as \cite{1978PASJ...30..671N,2004RvMP...76..125M} 
\begin{equation}
\mu = \frac{\mu_{\rmn{sphere}}}{\mu_{\rmn{critical}}}
\end{equation}
where the ratio between the magnetic and gravitational forces of the sphere is given by
\begin{equation}
\mu_{\rmn{sphere}} = \frac{M}{\pi{}r^{2}_{\rmn{sphere}}B_0}
\end{equation}
and the critical ratio (where the magnetic and gravitational force are in equilibrium) given by
\begin{equation}
\mu_{\rmn{critical}} = \frac{2c_1}{3} \sqrt{\frac{5}{\pi{}G\mu_0}} \text{~,}
\end{equation}
with $c_1 = 0.53$ \cite{1976ApJ...210..326M}. We use a mass-to-flux ratio of 5, which gives a plasma $\beta = 4.5$ inside the sphere. \diff{\ignore{The initial velocity profile of} The sphere is \ignore{equivalent to} set in solid-body rotation with }$\Omega = 1.77 \times 10^{-13} \text{~rad~s}^{-1}$\diff{ such that the magnitude of the ratio of rotational to gravitational energy is initially }$\approx 0.005$\diff{. The magnetic field and rotation axes as both aligned with the }$z$\diff{-axis, \textit{i.e.} the parameter }$\vartheta$\diff{ defined in LBP2015 as the angle between the rotation and field axes, is set to }$\vartheta = 0\degree$\diff{.}

We add sink particles when the density exceeds $10^{-10} \rmn{~g~cm}^{-3} \approx 10^7~\rho_0$. Previously, accretion radii smaller than 5 AU could not be used due to the instability discussed earlier (at $r_{\rmn{acc}} > 5 \rmn{~AU}$ the sink is so large that material is accreted before a sufficiently large density gradient can be created), here we use $r_{\rmn{acc}} = 1 \rmn{~AU}$ but we have also performed calculations with significantly smaller sink particles. 

As in \cref{sec:tests}, we have performed calculations with an unmodifed code (P0) each average (P1-4), and also with an arithmetic mean applied only to the induction equation or momentum equation (P1a and P1b).

Similar to \cref{sec:tests}, in \cref{fig:MultiP} we plot the column density of the collapsed molecular cloud core at $t = 25~670~\rmn{yrs}$ for models P0-4. The violent instability in the P0 model can be much more clearly seen here. \diff{We would expect the large cold sphere to collapse under its own gravity, and form a thin pseudo-disc around a dense core, ultimately producing a magnetically driven bipolar jet - as seen in P1 and in LBP2015.} In contrast to the earlier tests, the P1a and P1b models both initially appear to be stable\diff{. However, they}\ignore{until} eventually \diff{both} fail\ignore{ing} indicating that in this more complicated model both equations become unstable\diff{.} \ignore{and i}\diff{I}n \cref{fig:GrabCompP} we do not observe the clear separation between the increase in magnetic and kinetic energies seen in the test model. \diff{Unlike the test cylinder model, the quadratic mean in P4 does not evolve with significant differences in the density profile.}

All four averages produce first hydrostatic core jets with velocities of approximately $8~\rmn{km~s}^{-1}$, comparable to those obtained with a larger 5 AU sink particle in \cite{2012MNRAS.423L..45P}, albeit slightly faster due to the smaller sink radius \cite{2003MNRAS.339.1223P}. The evolution of these models can then be followed until the jet hits the edge of the periodic box \textit{ca.} 4,000 yrs later. No significant differences are seen in the evolution of each of these models, combined with the results from models D1-4 in the previous section this indicates that our average $h$ formalism is robust to the choice of average provided the full density gradient is sampled, however, in some low resolution situations the quadratic mean may be a poor choice. 

\section{Conclusion}
\label{sec:conc}
The ability to perform stable SPMHD calculations where steep density gradients are present has, even with the major advances in recent years, been impossible. Having developed a slight modification to the equations of SPMHD by using an average smoothing length term to fully sample these steep density gradients (\cref{sec:avh}), we were able in LBP2015 to model the collapse of a magnetised cloud core with a $\leq{} 1$ AU sink particle and with an arbitrary choice of field geometry. Hitherto, such work had been impossible as the calculations would violently explode due to (one or more of) a rapidly growing field or magnetic force caused by the equations being evaluated over extremely dissimilar neighbour sets. In this paper we then considered the effect the choice of average (\cref{sec:averages}) and which equations are modified on the numerical stability of the SPMHD equations. We find that except in low resolution tests the formalism is robust to the choice of average provided the full density gradient is sampled; however, owing to the results in our low resolution tests we recommend the use of the Pythagorean means only. Whilst in some test models only the SPMHD induction equation needs to be modified to ensure stability, in more complicated situations modifications to the anisotropic component of the momentum equation is also essential to maintain a stable simulation.


\section*{Acknowledgment}
This work was supported by the European Research Council under the European Community's Seventh Framework Programme (FP7/2007-2013 Grant Agreement No. 339248). BTL also acknowledges support from an STFC Studentship and Long Term Attachment grant.

MRB's visit to Monash was funded by an International Collaboration Award from the Australian Research Council (ARC) under the Discovery Project scheme grant DP130102078. DJP acknowledges funding from the ARC via DP130102078 and FT130100034.
STFC
The calculations for this paper were performed on the DiRAC Complexity machine, jointly funded by STFC and the Large Facilities Capital Fund of BIS, and the University of Exeter Supercomputer, a DiRAC Facility jointly funded by STFC, the Large Facilities Capital Fund of BIS and the University of Exeter.

Rendered plots were produced using the \texttt{SPLASH} \cite{2007PASA...24..159P} visualisation programme.



\bibliographystyle{IEEEtran.bst}
\bibliography{IEEEabrv,SPHERIC}

\begin{thebibliography}{10}
\providecommand{\url}[1]{#1}
\csname url@samestyle\endcsname
\providecommand{\newblock}{\relax}
\providecommand{\bibinfo}[2]{#2}
\providecommand{\BIBentrySTDinterwordspacing}{\spaceskip=0pt\relax}
\providecommand{\BIBentryALTinterwordstretchfactor}{4}
\providecommand{\BIBentryALTinterwordspacing}{\spaceskip=\fontdimen2\font plus
\BIBentryALTinterwordstretchfactor\fontdimen3\font minus
  \fontdimen4\font\relax}
\providecommand{\BIBforeignlanguage}[2]{{%
\expandafter\ifx\csname l@#1\endcsname\relax
\typeout{** WARNING: IEEEtran.bst: No hyphenation pattern has been}%
\typeout{** loaded for the language `#1'. Using the pattern for}%
\typeout{** the default language instead.}%
\else
\language=\csname l@#1\endcsname
\fi
#2}}
\providecommand{\BIBdecl}{\relax}
\BIBdecl

\bibitem{1977MNRAS.181..375G}
R.~A. {Gingold} and J.~J. {Monaghan}, ``{Smoothed particle hydrodynamics -
  Theory and application to non-spherical stars},'' \emph{\mnras}, vol. 181,
  pp. 375--389, Nov. 1977.

\bibitem{2004MNRAS.348..123P}
D.~J. {Price} and J.~J. {Monaghan}, ``{Smoothed Particle Magnetohydrodynamics -
  I. Algorithm and tests in one dimension},'' \emph{\mnras}, vol. 348, pp.
  123--138, Feb. 2004.

\bibitem{2001ApJ...561...82B}
S.~{B{\o}rve}, M.~{Omang}, and J.~{Trulsen}, ``{Regularized Smoothed Particle
  Hydrodynamics: A New Approach to Simulating Magnetohydrodynamic Shocks},''
  \emph{\apj}, vol. 561, pp. 82--93, Nov. 2001.

\bibitem{2012JCoPh.231.7214T}
T.~S. {Tricco} and D.~J. {Price}, ``{Constrained hyperbolic divergence cleaning
  for smoothed particle magnetohydrodynamics},'' \emph{Journal of Computational
  Physics}, vol. 231, pp. 7214--7236, Aug. 2012.

\bibitem{2012MNRAS.425.1068D}
W.~{Dehnen} and H.~{Aly}, ``{Improving convergence in smoothed particle
  hydrodynamics simulations without pairing instability},'' \emph{\mnras}, vol.
  425, pp. 1068--1082, Sep. 2012.

\bibitem{2005MNRAS.364..384P}
D.~J. {Price} and J.~J. {Monaghan}, ``{Smoothed Particle Magnetohydrodynamics -
  III. Multidimensional tests and the constraint},'' \emph{\mnras}, vol. 364,
  pp. 384--406, Dec. 2005.

\bibitem{1997JCoPh.136..298M}
J.~J. {Monaghan}, ``{SPH and Riemann Solvers},'' \emph{Journal of Computational
  Physics}, vol. 136, pp. 298--307, Sep. 1997.

\bibitem{1997JCoPh.136...41M}
J.~P. {Morris} and J.~J. {Monaghan}, ``{A Switch to Reduce SPH Viscosity},''
  \emph{Journal of Computational Physics}, vol. 136, pp. 41--50, Sep. 1997.

\bibitem{2013MNRAS.436.2810T}
T.~S. {Tricco} and D.~J. {Price}, ``{A switch to reduce resistivity in smoothed
  particle magnetohydrodynamics},'' \emph{\mnras}, vol. 436, pp. 2810--2817,
  Dec. 2013.

\bibitem{2007MNRAS.374.1347P}
D.~J. {Price} and J.~J. {Monaghan}, ``{An energy-conserving formalism for
  adaptive gravitational force softening in smoothed particle hydrodynamics and
  N-body codes},'' \emph{\mnras}, vol. 374, pp. 1347--1358, Feb. 2007.

\bibitem{1995JCoPh.116..123S}
J.~W. {Swegle}, D.~L. {Hicks}, and S.~W. {Attaway}, ``{Smoothed Particle
  Hydrodynamics Stability Analysis},'' \emph{Journal of Computational Physics},
  vol. 116, pp. 123--134, Jan. 1995.

\bibitem{2004MNRAS.348..139P}
D.~J. {Price} and J.~J. {Monaghan}, ``{Smoothed Particle Magnetohydrodynamics -
  II. Variational principles and variable smoothing-length terms},''
  \emph{\mnras}, vol. 348, pp. 139--152, Feb. 2004.

\bibitem{1995MNRAS.277..362B}
M.~R. {Bate}, I.~A. {Bonnell}, and N.~M. {Price}, ``{Modelling accretion in
  protobinary systems},'' \emph{\mnras}, vol. 277, pp. 362--376, Nov. 1995.

\bibitem{1997MNRAS.288.1060B}
M.~R. {Bate} and A.~{Burkert}, ``{Resolution requirements for smoothed particle
  hydrodynamics calculations with self-gravity},'' \emph{\mnras}, vol. 288, pp.
  1060--1072, Jul. 1997.

\bibitem{2008ApJ...676.1088M}
M.~N. {Machida}, S.-i. {Inutsuka}, and T.~{Matsumoto}, ``{High- and
  Low-Velocity Magnetized Outflows in the Star Formation Process in a
  Gravitationally Collapsing Cloud},'' \emph{\apj}, vol. 676, pp. 1088--1108,
  Apr. 2008.

\bibitem{1969MNRAS.145..271L}
R.~B. {Larson}, ``{Numerical calculations of the dynamics of collapsing
  proto-star},'' \emph{\mnras}, vol. 145, p. 271, 1969.

\bibitem{1978PASJ...30..671N}
T.~{Nakano} and T.~{Nakamura}, ``{Gravitational Instability of Magnetized
  Gaseous Disks 6},'' \emph{\pasj}, vol.~30, pp. 671--680, 1978.

\bibitem{2004RvMP...76..125M}
M.-M. {Mac Low} and R.~S. {Klessen}, ``{Control of star formation by supersonic
  turbulence},'' \emph{Reviews of Modern Physics}, vol.~76, pp. 125--194, Jan.
  2004.

\bibitem{1976ApJ...210..326M}
T.~C. {Mouschovias} and L.~{Spitzer}, Jr., ``{Note on the collapse of magnetic
  interstellar clouds},'' \emph{\apj}, vol. 210, p. 326, Dec. 1976.

\bibitem{2012MNRAS.423L..45P}
D.~J. {Price}, T.~S. {Tricco}, and M.~R. {Bate}, ``{Collimated jets from the
  first core},'' \emph{\mnras}, vol. 423, pp. L45--L49, Jun. 2012.

\bibitem{2003MNRAS.339.1223P}
D.~J. {Price}, J.~E. {Pringle}, and A.~R. {King}, ``{A comparison of the
  acceleration mechanisms in young stellar objects and active galactic nuclei
  jets},'' \emph{\mnras}, vol. 339, pp. 1223--1236, Mar. 2003.

\bibitem{2007PASA...24..159P}
D.~J. {Price}, ``{splash: An Interactive Visualisation Tool for Smoothed
  Particle Hydrodynamics Simulations},'' \emph{\pasa}, vol.~24, pp. 159--173,
  Oct. 2007.

\end{thebibliography}
%
%
%

\end{document}